\documentclass[twocolumn,showpacs,preprintnumbers,amsmath,amssymb,epsf]{revtex4}

\usepackage{epsf}

\def\fun#1#2{\lower3.6pt\vbox{\baselineskip0pt\lineskip.9pt
\ialign{$\mathsurround=0pt#1\hfill##\hfil$\crcr#2\crcr\sim\crcr}}}

\def\rn{}
\def\nn#1 #2{#2. #1}                            
\def\nnn#1 #2 #3{#2. #3. #1}                    
\def\nnnn#1 #2 #3 #4{#2. #3. #4. #1}            
\def\nnnnn#1 #2 #3 #4 #5{#2. #3. #4 #5. #1}     

\def\rf#1;#2;#3;#4;#5 {{\frenchspacing\par\rn#1, #3 {\bf #4}, #5 (#2). \par}}
\def\rfbook#1;#2;#3;#4;#5 {{\frenchspacing\par\rn#1, {\it #3} (#5, #4, #2).\par}}
\def\rfprep#1;#2;#3 {{\par\frenchspacing\rn#1, #3 (#2).\par}}

\newcommand{\nc}{\newcommand}

\nc{\be}[1]{\begin{equation}\mbox{$\label{#1}$}}
\nc{\bea}[1]{\begin{eqnarray} \mbox{$\label{#1}$}}
\nc{\Section}[2]{\section{#2}\label{#1}}
\nc{\Bibitem}[1]{\bibitem{#1}}
\nc{\Label}[1]{\label{#1}}

\nc{\Mpc}{Mpc/h}
\nc{\vev}[1]{\langle #1 \rangle}

\nc{\eea}{\end{eqnarray}}
\nc{\ee}{\end{equation}}

\def\lcdm{$\Lambda$CDM~}

\def\etal{{et al. }}

\def\ltsima{$\; \buildrel < \over \sim \;$}
\def\gtsima{$\; \buildrel > \over \sim \;$}
\def\simlt{\lower.5ex\hbox{\ltsima}}
\def\simgt{\lower.5ex\hbox{\gtsima}}
\nc{\w}{$w_2(\theta)$\ }
\nc{\ie}{i.e.} 
\nc{\eg}{e.g.}
\def\q{{\bf q} }

\begin{document}

\title{3-point temperature anisotropies in WMAP: \\
Limits on CMB non-Gaussianities and non-linearities}

\author{E.Gazta\~naga$^{1,2}$, J.Wagg$^2$}
\address{$^1$Institut d'Estudis Espacials de Catalunya, IEEC/CSIC, Gran
Capit\'an 2-4, 08034 Barcelona, Spain}
\address{$^2$Instituto Nacional de Astrof\'{\i}sica, \'Optica y Electr\'onica
(INAOE), Aptdo. Postal 51 y 216, Puebla, Mexico}






\begin{abstract}
  We present a study of the 3-pt angular correlation function 
  $w_3= <\delta_1\delta_2\delta_3>$ of
(adimensional) temperature anisotropies measured by the Wilkinson Microwave
Anisotropy Probe (WMAP). Results can be normalized to the 2-point function 
$w_2 =<\delta_1\delta_2>$
in terms of the hierarchical: $q_3 \sim w_3/w_2^2$ or dimensionless: $d_3 \sim
w_2/w_2^{3/2}$ amplitudes. Strongly non-Gaussian models are generically expected
to show $d_3 > 1$ or $q_3 > 10^3 d_3$. Unfortunately, this is comparable to the
cosmic variance on large angular scales. For Gaussian primordial models, $q_3$
gives a direct measure of the non-linear corrections to temperature
anisotropies in the sky: $\delta = \delta_L + f_{NLT} (\delta_L^2 - <\delta_L^2>)$
with $f_{NLT} = q_3/2$ for the leading order term in $w_2^2$. 
We find good agreement with the
Gaussian hypothesis $d_3 \sim 0$ within the cosmic variance of \lcdm
simulations (with or without a low quadrupole). The strongest constraints on $q_3$ come
from scales smaller than 1 degree. We find $q_3 =19 \pm 141$ for (pseudo)
collapsed configurations and an average of $q_3 = 336 \pm 218$ for non-collapsed
triangles. The corresponding non-linear coupling parameter, $f_{NL}$, for
curvature perturbations $\Phi$, in the Sachs-Wolfe (SW) regime is: $f_{NL}^{SW} =
q_3/6$, while on degree scales, the extra power in acoustic oscillations
produces $f_{NL} \sim q_3/30$ in the \lcdm model. Errors are dominated by cosmic
variance, but for the first time they begin to be small enough to constrain the
leading order non-linear effects with coupling of order unity.

\end{abstract}


\keywords{cosmology -- cosmic microwave background}

\maketitle


\section{Introduction}

We study  the cosmic microwave background (CMB)
anisotropies measured by WMAP \citep{Ben03} 
from the point of view of the 3-point angular correlation $w_3$. 
Current models for structure formation predict that the initial
seeds, that later give rise to the observed structure in the universe 
, were Gaussian. If so, one would expect the angular 3-point CMB temperature correlations 
to be almost zero. Even for a Gaussian field, any non-linear processing
of the primordial perturbations would lead to small non-Gaussianities.
For an excellent review see \citep{Kom03} and references
therein.

Angular correlations have some advantages over spherical harmonic analyses. 
The main disadvantage is that different angular bins
are highly correlated, which means that we need to use the
full covariance matrix to assess the significance of any departures
between measurements and models. The advantage of using
the N-point function is that correlations can be easily sampled
from any region of the sky, even when a foreground mask is required, provided
the region is large enough. In comparison, harmonic decomposition
breaks the angular symmetry in the sky into orthogonal bases which
have an arbitrary phase orientation. This is not a problem in the case
of full sky coverage, but the results will be coordinate-dependent if
a mask is used. Moreover, higher order correlations, such as the bispectrum, 
have to focus on some small portion of the mutidimensional 
space of triangle configurations (see \citep{Kom03}).
The formal equivalence between the bispectrum and the 3-point
function is hardly realized in practice and as such it is of interest
to study both separately.

Our analysis procedure is similar to that presented in 
\citep{Gaz03},
to which the reader can refer for further details. We use the same simulations
where we've generated maps using the standard \lcdm model and a low-Q \lcdm model, 
a fiducial variation of the  \lcdm model with low values of the quadrupole and 
octopole. We concentrate our analysis on the
V-band WMAP maps (FWHM = 21' at 61\,GHz) generated using 
HEALPix \citep{Gor99} with pixel resolution n=512 (7' pixels)
and to which we apply the 
conservative WMAP kp0 foreground mask \citep{Ben03}, The
resulting maps have approximately 2.4 million pixels. 
\section{Correlation functions from WMAP}


The 2-point angular correlation function is defined as the 
expectation value or mean cross-correlation of temperature fluctuations:
\be{delta}
\delta(\q) \equiv {\Delta T(\q)\over{T_0}}
\ee
at two positions $\q_1$ and $\q_2$ in the sky:
\be{w2def}
w_2(\theta) \equiv \vev{ \delta({\bf\q_1}) \delta({\bf\q_2}) },
\ee
where $\theta = |\bf{\q_2}-\bf{\q_1}|$, assuming that
the distribution is statistically isotropic. We normalize
fluctuations to be dimensionless, $T_0= 2.73$K, or dimensional
with $T_0=1$.
To estimate $w_2(\theta)$ from the pixel maps we use:
\be{w2est}
w_2(\theta) = {\sum_{i,j} \delta_i \delta_j ~w_i~w_j \over{\sum_{i,j} w_i~w_j}},
\ee
where $\delta_i$ and $\delta_j$ are the observed temperature
fluctuations on the sky
and the sum extends to all pairs $i,j$ separated by $\theta \pm \Delta\theta$. 
The mean temperature fluctuation is subtracted so that $\vev{\delta_i}=0$.
The weights $w_i$ can be used to minimize the variance when the pixel
noise is not uniform, however this introduces larger cosmic variance.
Here we follow the WMAP team and use uniform weights (i.e. $w_i=1$). 
At the largest scales we use bins $\Delta\theta$ whose size is proportional 
to the square root of  the angle $\theta$ and use linear bins of
$\Delta\theta \simeq 0.26$ degrees at the smallest scales.

\subsection{3-point function}

In a similar fashion, the 3-point angular correlation function 
is defined as the cross-correlation of temperature fluctuations 
at 3 positions $\q_1$, $\q_2$ and $\q_3$ on the sky:
\be{w3def}
w_3(\theta_{12}, \theta_{13}, \theta_{23}) 
\equiv \vev{ \delta({\bf\q_1}) \delta({\bf\q_2}) \delta({\bf\q_3}) },
\ee
where $\theta_{ij} = |\bf{\q_i}-\bf{\q_j}|$. 
Consider a triplet of pixels with labels 1,2,3 on the sky. Let 
$\theta_{12}$ and $\theta_{13}$ be the angular separations between the 
corresponding pairs of pixels and $\alpha$ the interior angle between 
these two sides of the triangle.
Assuming that
the distribution is statistically isotropic, $w_3$ only depends
on the sides of the triangles that define the 3 positions on the 
sky. One can characterize the 
configuration-dependence of the three-point function by 
studying the behavior of $w_3(\alpha)$ for fixed
$\theta_{12}$ and $\theta_{13}$.

To estimate $w_3(\theta_{12}, \theta_{13}, \theta_{23})$ 
from the pixel maps we use:
\be{w3est}
w_3(\theta_{12}, \theta_{13}, \theta_{23}) 
= {\sum_{i,j,k} \delta_i \delta_j \delta_k ~w_i~w_j~w_k \over{\sum_{i,j,k} w_i~w_j~w_k}},
\ee
where $\delta_i$, $\delta_j$ and $\delta_k$ are the temperature differences in the map
and the sum extends to all triplets separated by 
$\theta_{12} \pm \Delta\theta$, $\theta_{13} \pm \Delta\theta$ and 
$\theta_{23} \pm \Delta\theta$. We use same bins and weights as in the 2-point
function above.

The 3-point function can either be normalized to a dimensionless amplitude:

\be{d3}
d_3  \equiv {w_3
 \over{(w_{12}w_{23}w_{13})^{3/2}}}
\ee
where $w_{ij}=w_2(\theta_{ij})$, or a hierarchical amplitude:

\be{q3}
q_3  \equiv {w_3
 \over{w_{12}w_{23}+w_{12}w_{13}+w_{23}w_{13}}}.
\ee

\subsection{Errors}

The covariance matrix between different angular bins
can be calculated from the simulations:
\bea{eq:covar}
C_{ij} &\equiv& 
\langle\Delta w_2(\theta_i)~\Delta w_2(\theta_j)\rangle\nonumber\\
&=& {1\over{N}} \sum_{L=1}^{N} \Delta w_2^L(\theta_i) \Delta w_2^L(\theta_j),
\eea
where $\Delta w_2^L(\theta_i) \equiv w_2^L(\theta_i) - \widehat{w_2}(\theta_i)$.
Here $w_2^L(\theta_i)$ is the 2-point or 3-point
function measured in the $L$-th realization ($L=1\dots N$)  and $\widehat{w_2}(\theta_i)$ 
is the mean value for the $N$ realizations.
The case $i=j$ gives the error variance: $\sigma_i^2 \equiv C_{ii}$.

An estimate of the errors for correlations measured on the real
sky can be obtained using a variation of the jackknife error scheme.
 This has the potential advantage of producing
an error estimate which is model independent.
The accuracy of the jack-knife covariance have been tested
for both WMAP \citep{Gaz03} and the APM and SDSS survey
\citep{scranton/etal:2002}, \citep{gaztanaga:2002a}, \citep{gaztanaga:2002b}.
 In the estimation,
the sample is first divided into $M$ separate regions on the sky, 
each of equal area. The analysis
is then performed $M$ times, on each occasion removing
 a different region. These are called
the (jackknife) subsamples, which we label $k=1 \dots M$. The estimated
statistical covariance for $w_{2}$ at scales $\theta_i$ and $\theta_j$
is then given by:
\bea{covarjack}
C_{ij}
&=& {M-1\over{M}} \sum_{k=1}^{M} \Delta w_{2}^k(\theta_i) \Delta
w_{2}^k(\theta_j) \\
& &\Delta w_{2}^k(\theta_i) \equiv w_{2}^k(\theta_i) - \widetilde{w_{2}}(\theta_i)
\eea
where $w_{2}^k(\theta_i)$ is the measure 2 or 3-point function in the 
$k$-th subsample ($k=1
\dots M$)  and $\widetilde{w_{2}}(\theta_i)$ is the mean value for the $M$ subsamples.
The case $i=j$ gives the error variance. Note how, if we increase the number of
regions $M$, the jackknife subsamples are larger and each term in the sum is smaller. 
We typically take $M=8$ corresponding to a division of the sphere into 8 octants.
Note that the 3-point correlation function is in general a function of 3 variables. We will
fix two of these variables and show variations of $w_3$ or $q_3$ as a function
of a single variable, for which we estimate the covariance with the above prescription.

\subsection{The algorithm}

Estimation of correlations in configuration space is quite time consuming.
A brute force algorithm to estimate all pair separations in a map with
$N$ pixels takes $\simeq N^2$ operations. To do all triplets in Eq.[\ref{w3est}]
we need $\simeq N^3$ operations. In the healpix $n=512$ WMAP maps 
(7 arcmin pixel's) we have 
$N \simeq 2.4 \times 10^6$ (with the Kp0 mask galactic cut) so a brute
force analysis
requires $\simeq 10^{20}$ operations, well beyond current computing power.
We overcome this difficulty in two steps. For large scales, $\theta>1$ degree,
we reduce the resolution of the maps to $n=64$ (1 degree resolution)
by averaging fluctuations over nearby pixels. The 
resulting correlation function contains the same information at scales larger than  
the new resolution scale.  Now the number of pixels is reduced to $N = 3 \times 10^4$ and 
the number of operations to $\simeq 10^{13}$, which lies within current computer
power (Teraflops). For scales smaller than a few 
degrees we can reduce the number of operations
by precalculating the distance between neighboring pixels. The number of operations
is then reduced to $N\times M^2$, where $M<<N$ is the number of neighbors.
In our case we have $M \simeq 10^3$ for pixels around
$\theta < 4$ degrees, and thus $N\times M^2 \simeq 10^{13}$ as in the
case for the larger angles. In our implementation we have an overlap between the large and 
small scale estimations at around $2-4$ degrees. The agreement between both
estimates here is quite good for both $w_2$ and $w_3$. In the case of the 2-point
function we have compared both estimates in the whole range $2-180$ degrees
and find excellent agreement.

\subsection{Results}

\begin{figure} 
\epsfxsize 3.3in 
\epsfbox{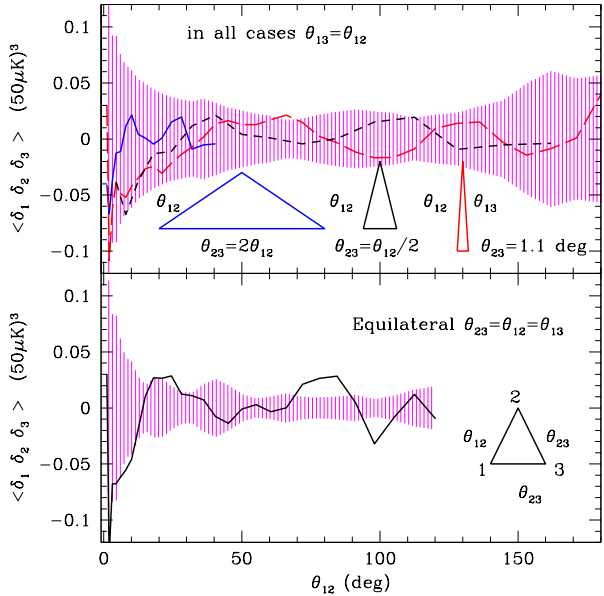}
\caption[Fig1] {\label{w3co} 
3-point function $w_3=< \delta_1 \delta_2 \delta_3>$
at large angular scales. Bottom spanel shows $w_3$ for 
equilateral triangles as a function of side $\theta_{12}$. The top panel 
shows results for isosceles triangles $\theta_{13}=\theta_{12}$
with $\theta_{23}=1.1deg$ (long dashed line),
$\theta_{23}=\theta_{12}/2$ (short dashed line) and
$\theta_{23}=2\theta_{12}$ (continuous line), in all cases
as a function of side $\theta_{12}$. 
Shaded regions are for the 68\% confidence levels estimated using jackknife errors.}
\end{figure}

\begin{figure} 
\epsfxsize 3.3in 
\epsfbox{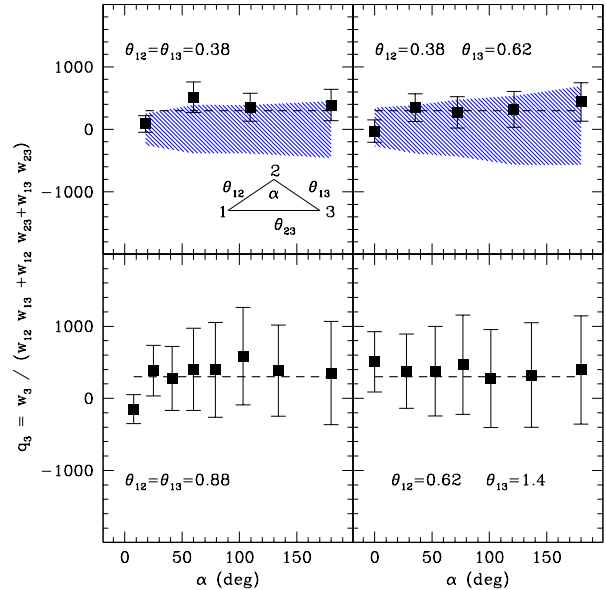}
\caption[Fig2] {\label{q3al} Reduced
 3-point function $q_3$
at small angular scales. Different panels show configurations
with triangles of fixed sides $\theta_{12}$ and $\theta_{13}$
(as labeled in each panel) and variable $\theta_{23}$ which
is given by the interior angle $\alpha$ between 
$\theta_{12}$ and $\theta_{13}$ ($\alpha=0$ corresponds
to $\theta_{23}=0$ and $\alpha=180$ corresponds
to $\theta_{23}=2 \theta_{12}$). Shaded regions are for the 68\%
confidence levels estimated using low-Q \lcdm simulations. Errorbars are from
jackknife variance.}
\end{figure}

Results are presented in Figures \ref{w3co} and \ref{q3al}. 
At large scales Fig.\ref{w3co}
shows the results of $w_3$ in units of $(50 \mu K)^3$ (eg $T_0=50 \mu K$
in Eq.[\ref{delta}]). Note that
$w_2^{1/2} \simeq 50 \mu K$ at small scales, so that $w_3 \simeq d_3 \equiv
w_3/w_2^{3/2}$  in these units (see Eq.[\ref{d3}]).
Errorbars in both cases are from jackknife subsamples and agree
well with errors estimated from the simulations (see Gazta\~naga etal 2003 for more details). 
The results for the equilateral and collapsed triangle configurations
can be compared to the COBE analysis presented in Fig.1 of \citep{Kog96}.
The pseudo-collapsed case in  \citep{Kog96} would
correspond here to $\theta_{23}=1.1deg$ here.
We can see how the COBE and WMAP results are in good agreement where
the two curves even see the very same bumps and valleys. This is not totally surprising given the
very good match between COBE and WMAP temperatures maps, pixel by pixel,
when smoothed on large scales. The top  panel in Figure \ref{w3co}
also shows two new triangle configurations: one that is elongated,
$\theta_{23}=\theta_{12}/2$ (short dashed line), and a wide triangle,
$\theta_{23}=2 \theta_{12}$ (continuous line). The results for intermediate
angular configurations are similar. In all cases the agreement with the Gaussian prediction
$w_3=0$ is excellent, given the strong covariance.

At small scales we show the 3-point function in terms of
the reduced amplitude $q_3$ in Eq.[\ref{q3}] where the best constraints
are found on the smallest scales. As can be seen in the figure,
as we approach triangles with side $\theta \simeq 1$degree, the cosmic errors
become large $\Delta q_3 \simeq 1000$. We have fit the data to
a constant value of $q_3$, independent of scale using a $\chi^2$-test:
\be{eq:chi}
\chi^2 = \sum_{i,j=1}^{N} \Delta_i ~ C_{ij}^{-1} ~ \Delta_j,
\ee
where $\Delta_i \equiv q_3^O(\alpha_i) - q_3^{fit}$
is the difference between the "observation" $O$ and the
$fitted$ value. In order to eliminate the degeneracies in the covariance matrix, we 
perform a Singular Value Decomposition (SVD) of the covariance matrix
(see Gazta\~naga \etal 2003 for more details).
A joint fit with the jackknife covariance matrix for collapsed configurations
with $\theta< 1$ degree, yields:
\be{q3cfit}
q_3^{collapsed} = 19 \pm 141. 
\ee
The best signal-to-noise is found for the
non-collapsed  triangle configurations with $\theta \simeq  0.5$ degree
we find:
\be{q3fit}
q_3 = 336 \pm 218  ~~~~~~~;~~~~  \theta= 0.5 ~deg
\ee
providing a marginal detection at about  90\% confidence (shown as dashed line in Fig.\ref{q3al}). This fit corresponds
to the jackknife covariance matrix. A similar result is found for
the covariance matrix in the low-q \lcdm simulations. The errors increase
by almost a factor of two when using the covariance of the standard
\lcdm simulations. This increase is due to the larger cosmic variance introduced
by the quadrupole and octopole which are larger than in the WMAP data
(see \citep{Spe03}, citep{Gaz03} for a discussion).

\section{Discussion}

The above results for the 3-pt function confirms previous analyses
(\citep{Kom03} and references therein) 
that find WMAP to be in good agreement with the Gaussian
hypothesis. Strong non-Gaussian statistics
with $d_3$ in Eq.[\ref{d3}] of order unity or larger are
ruled out. On larger (COBE) angular scales the results are
dominated by cosmic variance so that $\Delta d_3 \simeq 1$ \citep{Sre93}
and $\Delta q_3 \simeq 1000$. On subdegree scales errors are smaller
and we find evidence for a marginal detection (90\% level) of 
non-Gaussianity in Eq.[\ref{q3fit}]. This result can be taken
as a detection or, at a higher significance level, a bound on possible non-linear 
effects. Expanding the observed temperature fluctuation $\delta$
as a perturbation over some primordial Gaussian 
linear contribution $\delta_L$:
\be{nonlinear}
\delta = \delta_L + f_{NLT}~ (\delta_L^2 - <\delta_L^2>),
\ee
we then have that to leading order in  $w_2$:
\be{w3h}
w_3 = 2 ~f_{NLT}~ (w_{12}w_{23}+w_{12}w_{13}+w_{23}w_{13}) + {\cal O}[w_2^4]
\ee
so that $q_3 \simeq 2 ~f_{NLT}$. 
The corresponding non-linear coupling parameter, $f_{NL}$,
for  curvature perturbations $\Phi$ is: 
\be{fnl}
\Phi = \Phi_L + f_{NL} ~(\Phi_L^2 - <\Phi_L^2>).
\ee
In the Sachs-Wolfe (SW) regime ($\delta \sim \Phi/3$) 
we have: $f_{NL}^{SW} \simeq q_3/6$, while on degree scales,
the extra power in acoustic oscillations produce  
$f_{NL} \simeq q_3/30$ in the case of \lcdm \cite{KS01}. This corresponds
to $f_{NL} \simeq 11.2 \pm 7.3$ for Eq.[\ref{q3fit}]
and  $f_{NL} \simeq 0.6 \pm 4.7$ for the collapsed case
Eq.[\ref{q3cfit}].  For a comparison with recent predictions see
\citep{Kom03,BU02,Acq02, Mal02, Zal03} and references therein.

Our  bounds on $f_{NL}$ seem marginally better than 
the ones presented in  \citep{Kom03} from the WMAP bispectrum,
but a direct comparison is not straight forward. First of all, our modelling provides
a measurement of non-linearities in the final
$\Delta T$ while \citep{Kom03} provide a modelling 
for non-linear primordial curvature perturbations $\Phi$, before
radiation transfer. Thus while limits on $q_3$ are model independent,
direct limits on $f_{NL}$ depend on the assumed model for radiation transfer.
Secondly, the bispectrum configurations used by \citep{Kom03}
are a subset of all possible triangle configurations so that both
approaches are neither equivalent nor  optimized in the same way.
In both cases, errors are dominated by cosmic variance, 
and they begin to be small enough to constrain
the leading order non-linear effects with coupling $f_{NL}$
which are of order unity.

\noindent
\begin{acknowledgements}  
EG acknowledged support from INAOE, the Spanish Ministerio de Ciencia y
Tecnologia, project AYA2002-00850, EC-FEDER funding and
the supercomputing center in Barcelona (CEPBA and CESCA/C4), where
part of these calculations were done. JW thanks Eric Hivon
for providing assistance during the installation of HEALPix
and acknowledges support from a CONACYT grant (39953-F) and a
scholarship from INAOE.
\end{acknowledgements}



\end{document}